\documentclass[aps,prl,reprint,twocolumn,showpacs,titlepage,10pt,longbibliography]{revtex4-1}

\usepackage{amssymb} 
\usepackage{amsmath} 
\usepackage[pdftex]{graphicx}
\usepackage{hyperref}
\hypersetup{colorlinks, linkcolor={blue},citecolor={blue}, urlcolor={blue}}
\usepackage{capt-of}
\usepackage{lipsum}
\usepackage[latin1]{inputenc}
\usepackage{color}

\newcommand{\abs}[1]{\left\lvert #1 \right\lvert }
 \newcommand{\modul}[1]{\left\lvert #1 \right\lvert }
\newcommand{\norme}[1]{\lvert \lvert #1 \lvert \lvert }
 \newcommand{\modulc}[1]{\left\lvert #1 \right\lvert ^{2}}

\newcommand{\av}[1]{\langle #1 \rangle}

\renewcommand{\vr}{\vec{r}}

\newcommand{\vn}{\hat{n}}

\newcommand{\E}{\vec{E}}

\newcommand{\ecdf}{\emph{ECDF}}

\definecolor{orange}{rgb}{0.99,0.69,0.07}
\definecolor{lgray}{gray}{0.75}
\definecolor{dgray}{gray}{0.15}
\begin{document}
\title{Wave Chaos in a Cavity of Regular Geometry with  Tunable Boundaries}

\author{Jean-Baptiste~Gros}
\thanks{jean-baptiste.gros@espci.fr}
\affiliation{Institut Langevin, CNRS UMR 7587, ESPCI Paris, PSL Research University, 75005 Paris, France}
\author{Philipp~del~Hougne}
\thanks{Current address:  Institut de Physique de Nice, CNRS UMR 7010, Universit\'{e} C\^{o}te d'Azur,
06100 Nice, France.}
\affiliation{Institut Langevin, CNRS UMR 7587, ESPCI Paris, PSL Research University, 75005 Paris, France}
\author{Geoffroy~Lerosey}
\affiliation{Greenerwave, ESPCI Paris Incubator PC'up, 75005 Paris, France}

\begin{abstract}
Wave chaotic systems underpin a wide range of research activities, from fundamental studies of quantum chaos via electromagnetic compatibility up to more recently emerging applications like microwave imaging for security screening, antenna characterisation or wave-based analog computation. To implement a wave chaotic system experimentally, traditionally cavities of elaborate geometries (bowtie shapes, truncated circles, parallelepipeds with hemispheres) are employed because the geometry dictates the wave field's characteristics. Here, we propose and experimentally verify a radically different paradigm: a cavity of regular geometry but with tunable boundary conditions, experimentally implemented  by leveraging a reconfigurable metasurface. Our results set new foundations for the use and the study of chaos in wave physics.\end{abstract}
\maketitle

For decades, wave chaos has been an attractive field of fundamental research concerning a wide variety of physical systems such as quantum physics~\cite{Houches1989,verb85,Berry1977,Seligman1993}, room or ocean acoustics~\cite{Mortessagne1993,Auregan2016,Tomsovic2010}, elastodynamics~\cite{Lobkis2000}, guided-wave optics~\cite{Pre_Doya} or microwave cavities~\cite{Stockmann1990,Deus1995,Kuhl2013,Dietz2015,Barthelemy2005}. The success of wave chaos is mainly due to its ability to describe such a variety of complex systems through a unique formalism which allows to derive a universal statistical behavior.
Indeed, since the Bohigas-Giannoni-Schmit conjecture~\cite{BGS} concerning the universality of level fluctuations in chaotic quantum spectra, it has become customary to analyse spectral and spatial statistics of wave systems whose ray counterpart is chaotic with the help of statistical tools introduced by random matrix theory (RMT) \cite{stoeckmann1999quantum,Guhr1998,Sokolov1989,rotter2009non,Kuhl2013,Gros2016}. 
In recent years, electromagnetic (EM) chaotic cavities have been involved in a variety of applications ranging from reverberation chambers for electromagnetic compatibility (EMC) tests \cite{Gros2015aem,Gros2014Wamot,Gradoni2014,Bastianelli2017,Arnaut2001,Sarrazin2017,Orjubin2009}, via  wavefront shaping \cite{Kaina2015,Dupre2015,del2016spatiotemporal} and microwave imaging \cite{Sleasman2016,TondoYoya2017,asefi2017use,yoya2018reconfigurable}, to applications in telecommunication and energy harvesting \cite{DelHougne201_harv,mimo}, indoor sensing \cite{DelHougne2018_MoDet,localiz}, antenna characterization \cite{davy2} and wave-based analog computation \cite{wbac}. All of these applications have in common to leverage the field ergodicity \footnote{The field ergodicity means that fields in chaotic systems are statistically equivalent to an appropriate random superposition of plane waves leading to a field which is statistically uniform,depolarized, and isotropic i.e the fields are speckle-like \cite{Gros2016,Gros2015,Pnini1996,Dorr1998,Kim2005,hemmady2005universal}.} of responses and eigenfields of chaotic cavities \cite{Gros2016,Gros2014Wamot}.
Traditionally, whether they are used to study fundamental physics or for applications, these cavities are associated with irregular geometries. They are often built from a parallelepipedic cavity  by modifying its geometry (for instance, by adding spherical caps or hemispheres~\cite{Deus1995,Gros2014Wamot,Gros2015aem,Kuhl2017,Bastianelli2017}) so that  its spatial and/or spectral statistics follow the universal RMT predictions~\cite{Gros2014Wamot}. Furthermore, most of these cavities include mechanical movable elements, so-called stirrers, adding to the chaoticity and allowing one to perform ensemble averaging (mode stirring)~\cite{Serra2017,hill2009electromagnetic}. 

In this Letter, we investigate a completely different approach to build a chaotic cavity, by \textit{only} modulating locally the boundary conditions of a cavity of completely regular geometry. Experimentally the tuning of the boundary conditions is achieved with a reconfigurable metasurface that covers parts of the cavity walls. First, we  study the amount of metasurface elements required to turn a regular cavity into a chaotic one. Since the metasurface is built  upon resonant elements, we consider frequencies matching  their operation band. The chaoticity of the cavity is evaluated by comparing the \textit{experimentally} observed  wave field  distribution  with RMT predictions for wave chaotic systems. The latter depend on a single experimentally evaluable parameter: the mean modal overlap $d$ \cite{Gros2016,Gros2015}. This overlap is defined at the operating frequency $f$ as the product of the average modal bandwidth $\Gamma_f$ and the mean density of states $n_f$. Second, by using an unexpected  efficiency of the metasurfaces outside  their operation band,  we show the  effectiveness of our approach irrespectively of the modal overlap regimes, the latter being a key parameter of all wave systems \cite{Gros2016,Cozza2011,Schroeder1962,Dupre2015}.

\begin{figure}[h]
	\includegraphics[width=\columnwidth]{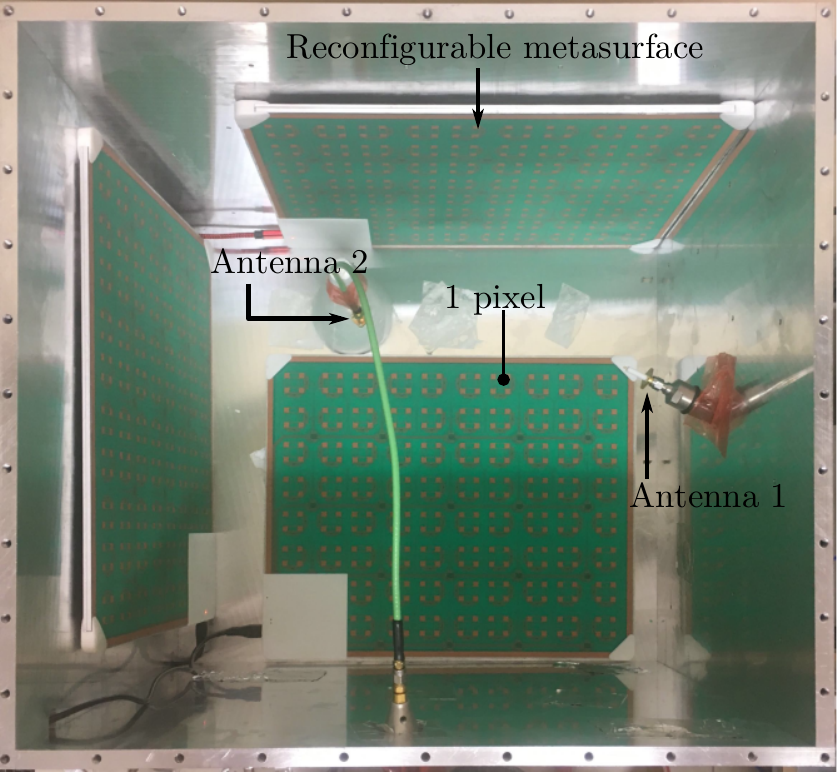}
	\caption{\label{fig:photo} Top view of the metallic parallelepipedic cavity ($42 \times  38.5  \times  35 \  \mathrm{cm}^3 $). Three walls are covered by reconfigurable metasurfaces ($76$ pixels per metasurface). Each metasurface pixel can be configured electronically to emulate a perfect electric or magnetic conductor. The wave field is probed by measuring the transmission between two antennas with a vector network analyzer (VNA). VNA and the cavity's top plate are not shown in this figure.}
\end{figure}

For our experiments we cover three contiguous and non-parallel walls of a metallic parallelepipedic cavity ($42 \times  38.5  \times  35 \  \mathrm{cm}^3 $) with electronically  reconfigurable metasurfaces (ERMs) \footnote{The ERMs are placed on three contiguous and non-parallel cavity walls to avoid potentially non-generic modes. \cite{Gros2014Wamot}}, without significantly altering the cavity geometry (see Fig.~\ref{fig:photo}). Each of the three metasurfaces consists of $76$ phase-binary pixels. The underlying working principle of hybridizing two resonances is outlined in Ref.~\cite{kaina2014hybridized}. By controlling the bias voltage of a PIN diode, each pixel can individually be configured to emulate the behavior of a quasi perfect electric or quasi perfect magnetic conductor. Stated differently, the phase of the tangential component of the field reflected by the pixel can be shifted by $\pi$. Note that our proposal to locally modulate the cavity's boundary conditions could also be implemented with other designs of tunable impedance surfaces, such as mushroom structures~\cite{sievenpiper2003two,sihvola2007metamaterials,Sleasman2016,li2018metasurfaces}. Since the design of our metasurface leverages resonant effects, the band of frequencies over which it displays the desired effect is a priori inherently limited. The ERM prototype  we use for our experiments has been designed to work efficiently within a $1\ \mathrm{GHz}$ bandwidth around $5.2\ \mathrm{GHz}$. 
 
To evaluate whether boundary condition modulations induced by ERMs are able to create a chaotic cavity, we compare the statistical distribution of the normalized intensity $I$ of Cartesian field components measured for ensembles configurations of ERMs with the theoretical RMT distribution. The main steps leading to the RMT distribution of the normalized field intensity of an ensemble of responses resulting from stirring are given in the Supplemental Material (interested readers are referred to \cite{Gros2016,Gros2015,CRC_review} for details). We recall here only the final RMT prediction which reads 
\begin{align}
P_{I;d}(I)&=\int_0^1    P^W_\rho (\rho) P(I;\rho) d\rho \label{eq:P:ensemble:rep}
\end{align}
where $\rho$ is the phase rigidity,
\begin{equation}
P(I;\rho)=\frac{1}{\sqrt{1-\modulc{\rho}}} \exp\left[ -\frac{I}{1-\modulc{\rho}} \right] \textrm{I}_0\left[\frac{\modul{\rho} I}{1-\modulc{\rho}} \right]\  \label{eq:Pnini}
\end{equation}
is the Pnini and Shapiro distribution \cite{Pnini1996,Gros2016,Kim2005}  and $ P^W_\rho$ is the phase rigidity distribution depending only on the mean modal overlap $d$  \cite{Gros2016,Gros2015} (see Supplemental Material for an analytical expression). For a 3D electromagnetic cavity of volume $V$, the mean density of states can be estimated with Weyl's law, which reads at leading order~\footnote{In Weyl's Law applied to a 3D EM cavity, there is no term proportional to $f$ and the total surface $S$ of the cavity; hence changing the configuration of an ERM is not expected to affect the cavity's mean density of states.}
\begin{equation}
n_f \simeq n_w(f)=\frac{8 \pi V }{c^3}f^2, \label{eq:nweyl}
\end{equation}  
where $c$ is  the speed of light and $f$ the mean of the considered frequency window. The mean modal overlap $d$ is thus related to $f$, $V$, the modal width $\Gamma_f$ and the composite quality factor $Q=f/\Gamma_f$ through 
\begin{equation}
d= n_f \Gamma_f\simeq\frac{8\pi V}{c^3Q} f^3 .  \label{eq:d}
\end{equation}

First, we are interested in the minimum number of metasurface pixels that have to emulate a perfect magnetic conductor to transform a regular metallic cavity into a chaotic one. To that end, we choose 500 random configurations of the three ERMs for which the overall number of PMC-like ('activated') pixels, $n_a$, is fixed and the $228-n_a$ remaining pixels are let in their PEC-like state (not 'activated'). 
For each configuration, we measure with a HP 8720D vector network analyzer the $S$-parameters between two monopole antennas for 1601 frequency points in a frequency window of $250$ MHz around $5.2$\,GHz where the pixels are the most efficient. This experiment is repeated for different value of $n_a  \in  
[2,122]$.
Then, for each set of experiments with fixed $n_a$, we extract for both antennas their  frequency-dependent coupling constants $\kappa_i(f)$ which read\cite{Kuhl2013,Kober2010,Kuhl2017,Fyodorov2005}:
\begin{equation} 
\kappa_i= \frac{\abs{1-\av{S_{ii}}}^2}{1-\abs{\av{S_{ii}}}^2}  \quad (i = 1, 2) , \label{eq:kappa}
\end{equation}  
where $S_{ii}(f)$ ($i=1,2$) are the reflection parameters and $\av{\cdot}$ denotes an ensemble average over random ERM configurations. Then, we deduce from the measurement of the transmission parameter $S_{12}(f)$ the normalized value of the amplitude of the Cartesian component of the electric field along the orientation of the monopole antenna 2 inside the cavity as \cite{Gros2015}
\begin{equation}
E_a=\E(\vr_2, f)\cdot \vn_a = \frac{S_{12}(f)}{\kappa_1\kappa_2}\label{eq:E:2:S12},
\end{equation}
where $\E(\vr_2, f)$ is the electric field at the position of antenna 2 and $\vn_a$  is the unit vector along the polarization of antenna 2. 
The RMT prediction in Eq.~\ref{eq:P:ensemble:rep} assumes that $\av{E_a}$ is vanishing. Physically, this means 
that static contributions such as direct processes (short path) are negligible \cite{Baranger1996,Hart2009,Yeh2010,yeh2010universal,Dietz2010}. Reasons for the presence of static contributions include directivity and relative positions of the antennas, as well as the ERMs' stirring efficiency. 
To extract the universal properties from our experiments that can be compared with RMT predictions, we numerically suppress the non-universal static contribution via the commonly used transformation $E_a\rightarrow E_a-\av{E_a}$ \cite{Dietz2010,Gradoni2014}, where $\av{\cdots}$ denotes averaging over ERM configurations \footnote{ In practical applications involving reconfigurable wave chaos, such a differential approach to remove the static contribution is routinely used to efficiently use the available degrees of freedom \cite{SMM_TM,localiz,wbac,yoya2018reconfigurable}.}. The universal and non-universal contributions in our data are discussed and displayed in detail in the Supplemental Material.%

For each set of experiments with fixed $n_a$, we compare the empirical cumulative distribution function (\textit{ECDF}) of the normalized field intensity $I=\abs{E_a}^2/\av{\abs{E_a^2}}$ of the ensemble of cavity configurations, $F_{n_a}(I)$, with the theoretical cumulative distribution function  
\begin{equation}
F_{I;d}(I)= \int_0^I P_{I;d}(x) dx  \label{eq:cdf},
\end{equation}
where we use the experimentally obtained value of $d$. To estimate $d$ with Eq.~\ref{eq:d}, we extract from our data the cavity's composite $Q$-factor as $Q=f/\Gamma_f=2\pi \tau f$, where $\tau=(2\pi\Gamma_f)^{-1}$ is the intensity decay time of the inverse Fourier transformed transmission signal $ \abs{\textrm{FT}(S_{21})}^2\propto \exp(-t/\tau)$. Around $5.2$~GHz, we thereby estimate $d=19.81$.
\begin{figure}[t!]
	\includegraphics[width=0.99 \columnwidth]{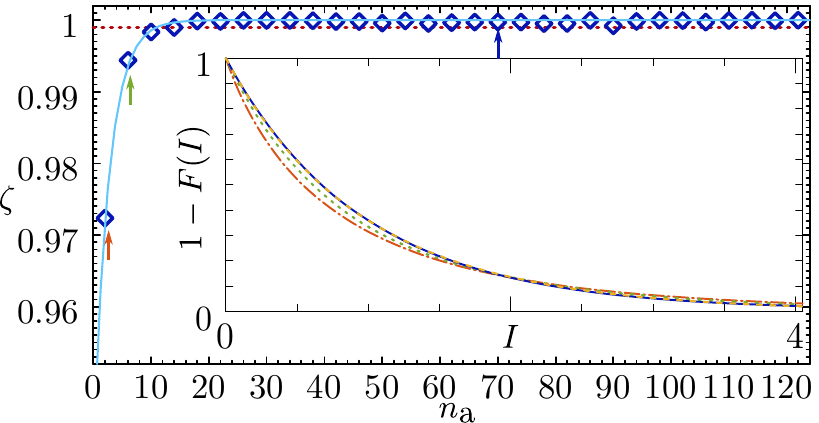}
	\caption{\label{fig:tran} Transition to chaotic behavior as the number of pixels emulating a perfect magnetic conductor, $n_a$, increases. Main plot:  $\diamond$, blue continuous curve and red doted line show, respectively, the experimental values of $\zeta(n_a)$ (see text for details), the  interpolation of $\zeta(n_a)$ by the heuristic function $f(x)=1- 0.06\exp(-0.373 x)$ and the limit $\zeta=0.999$ above which $F_{n_a}(I)$ is in good agreement with $F_{I;d}(I)$. Inset: red dashed-dotted line, green dotted line, blue continuous lines and orange dashed line correspond, respectively, to $F_2(I)$, $F_6(I)$, $F_{70}(I)$ and $F_{I;d}(I)$. Arrows in main plot locate the associated values of $\zeta(n_a)$. }
\end{figure} 
The deviation of the measured \ecdf\,of field intensity $F_{n_a}(I)$ from the RMT prediction $F_{I;d}(I)$ with $d=19.81$  is then estimated via the parameter $\zeta(n_a)$ defined as
\begin{equation}
\zeta(n_a)=1-\av{(F_{n_a}-F_{I;d})^2}_I/\av{(F_{n_a}-\av{F_{n_a})^2}}_I . \label{eq:zeta}
\end{equation}
In Fig~\ref{fig:tran}, we present the results. The diamonds ($\diamond$) correspond to the experimentally obtained values of $\zeta(n_a)$. A good agreement between the empirical $F_{n_a}(I)$ and the RMT prediction $F_{I;d}(I)$ is guaranteed as soon as $\zeta(n_a)\geq0.999$. This is illustrated in the inset of Fig~\ref{fig:tran} with the \emph{ECDFs}  $F_2(I)$,  $F_6(I)$,  $F_{70}(I)$ corresponding respectively to cases of $\zeta<0.99$, $0.99\leq \zeta<0.999$ and $\zeta \geq0.999$. Among  the three \emph{ECDFs}  shown, only $F_{70}(I)$ corresponding to the   case $\zeta \geq0.999$,   is in good agreement with the RMT prediction $F_{I;d}(I)$. Finally, to estimate the minimum number of activated pixels, $n_\textrm{min}$, required to obtain a chaotic cavity, we interpolate the measured $\zeta(n_a)$ by a heuristic function $f(x)=1-a\exp(-c x)$ and search the value $x_\textrm{min}$ such that $f(x_\textrm{min})=0.999$. The fit yields $a=0.06 \pm 3.1\% $, $c=0.373\pm 3.4\%$ and $x_\textrm{min}\simeq 10.98$. Therefore, in the considered cavity, $n_\textrm{min}=11$. This number depends obviously on the utilized metasurface design.

\begin{table}[b!]
	\caption{\label{tab1} Identification of five frequency windows, a) to e), with different mean modal overlaps $d$ (see Eq.~\ref{eq:d}). The central frequency $f$, the experimentally evaluated composite $Q$-factor and the mode number $N_w$ of the cavity at $f$ are indicated.}
\begin{tabular}{|c|c|c|c|c|}
	\hline
	\quad label   \quad &   \quad \quad \quad $f\ /\ \textrm{GHz}$ \quad  \quad \quad&    \quad  $Q$  \quad  \quad  &  \quad  $d$  \quad \quad   & \quad  \quad $N_w(f)$ \quad  \quad   \\
	\hline
	a) &  $1.84$ & $813$ & $0.4$ &$108$ \\
	\hline
	b) &  $3.1$ & $721$ & $1.98$ &$525$ \\
	\hline
	c) &  $3.6$  & $717$ & $3.47$ &$822$ \\
	\hline
	d) &  $4.5$& $747$ & $6.45$ &$1606$ \\
	\hline
	e) &  $5.2$ & $375$ & $19.81$& $2479$ \\
	\hline
\end{tabular} 
\end{table}

\begin{figure}[t!]
	\includegraphics[width=\columnwidth]{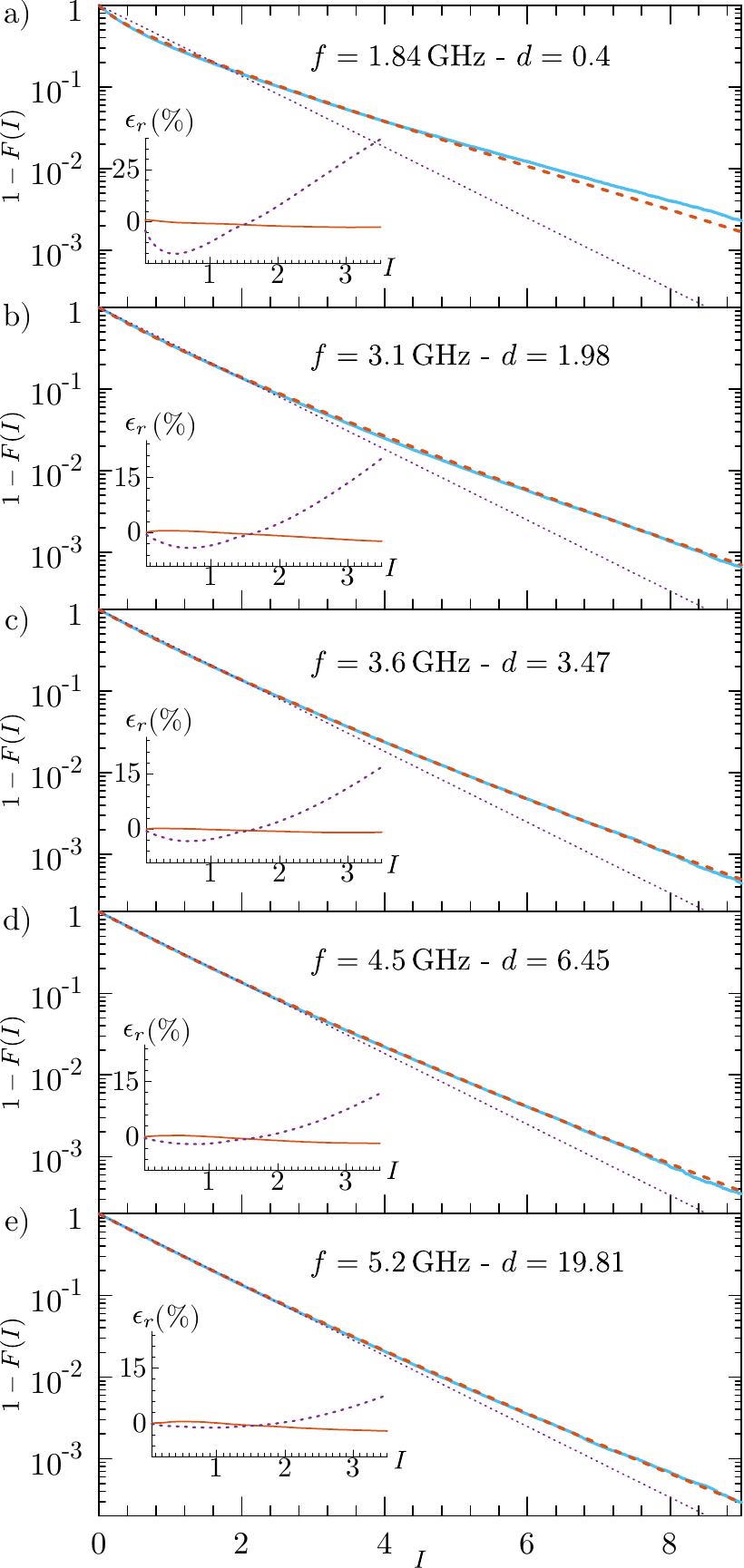}
	\caption{\label{fig:spa}Comparison of the experimentally observed wave fields with the behavior expected in a wave chaotic system, using as metric  the measured \ecdf\ of normalized field intensity $F(I)$. For five frequency windows a)-e) with different modal overlap $d$, the continuous blue and red dashed lines correspond, respectively, to $F(I)$ and the RMT prediction $F_{I;d}(I)$ (see Eq.~\ref{eq:cdf}). For reference, the cumulative distribution of the Hill-Ericson-Schroeder regime is also indicated (dotted purple). Insets show the relative errors $\epsilon_r$ between the \ecdf\ and the RMT prediction (continuous red), as well as the Hill-Ericson-Schroeder regime (dotted purple).}
\end{figure}

Having demonstrated that in a regular cavity equipped with ERMs chaotic behavior can be observed within the ERMs' operation band, we now consider frequencies outside this band allowing us to explore different regimes of modal overlap. Indeed, although the ERM pixels are individually less efficient far outside their designed operating band (the phase difference between the two states is well below $\pi$), surprisingly we observe that collectively they are still able to sufficiently alter the boundary conditions to create wave-chaotic behavior. Hence we now choose 9000 fully random configurations of the 228 pixels. For each ERM configuration, we measure the $S$-parameters between the monopole antennas for 1601 frequency points in $\left[ 1.8\mathrm{ GHz}, 5.8 \mathrm{ GHz}   \right] $. At this point, we draw the reader's attention to the fact that most of RMT predictions assume that the mean density of state, the coupling strength of antenna,  the absorption and the  ensuing mean modal overlap are constant\cite{stoeckmann1999quantum,Sokolov1989,Kuhl2013,Kober2010,Savin2017,Gradoni2014,Gros2014Wamot,Gros2015,Kuhl2017,Fyodorov2012,Gros2014,Poli2012,Kumar2013,Dietz2010,Guhr1998}. Practically, this means that we assume these quantities to vary only slightly within the investigated frequency range. Obviously, in the present study none of the above mentioned parameters are slightly varying on the full frequency range from $1.8$\,GHz to $5.8$ \,GHz, especially the mean density of state. Therefore, we focus our study on a subset of five frequency windows of 150 MHz width, labeled a) to e) and respectively centered on $1.84$ \,GHz, $3.1$ \,GH., $3.6$ \,GHz, $4.5$ \,GHz and $5.2$ \,GHz. Table~\ref{tab1} indicates for each of these frequency windows the estimated composite $Q$-factor, the associated value of the modal overlap and the mode number of the cavity, given by $N_w(f)=\int_0^f n_w(x) dx$. Then, we can study the field intensity distribution of the ensemble of cavity configurations, and hence the chaoticity, for different modal overlap regimes ranging from low modal overlap ($d<1$) around the 108$^{th}$ mode of the cavity to very high modal overlap ($d\simeq 20 \gg 1$) around its 2479$^{th}$ mode. For each frequency window in Table~\ref*{tab1}, we compare as before the measured \ecdf\,of the normalized  field intensity with the theoretical cumulative distribution function $F_{I;d}$, given by Eq.~\ref{eq:cdf}, using the corresponding experimentally measured value of $d$.
The results are shown in Fig.~\ref{fig:spa} where the panels a)-e) correspond to frequency windows a)-e) in Table~\ref{tab1}. In each panel of Fig.~\ref{fig:spa}, the continuous blue curve, the dashed red curve and the purple dotted curve correspond, respectively, to the complementary \ecdf,  $1-F(I)$, of experimental data, the RMT prediction $1-F_{I;d}(I)$ (equation~\ref{eq:cdf}) with $d$ given in Table~\ref{tab1}, and the complementary cumulative distribution function for the Hill-Ericson-Schroeder regime \cite{Lemoine2007b,hill2009electromagnetic,Dietz2010,Mortessagne1993}. The latter corresponds to the limit of very high modal overlap. From the very low modal overlap  regime with $d=0.4$ around the $108^{th}$ mode of the cavity (Fig~\ref{fig:spa}.a)) to the very high modal overlap regime with $d=19.81$ around the $2479^{th}$ mode of the cavity (Fig~\ref{fig:spa}.e)), we observe a very good agreement over three decades between the \textit{ECDF} of the normalized  field intensity of the ensemble of cavity's configurations and the RMT prediction for chaotic cavities. Hence, the cavity in Fig.~\ref{fig:photo} displays the universal statistical  behavior expected in chaotic cavities  when we randomly modulate its boundary conditions.

In the EMC community, the idea to use an electronically reconfigurable reverberation chamber to stir the EM field was previously proposed \cite{Kingler2005,Serra2017}, but had not been experimentally demonstrated to date. More recently, it was proposed to used a  metasurface to improve the field uniformity in a reverberation chamber \cite{Wanderlinder2017}. The idea of improving field uniformity is closely related to that of making the cavity chaotic \cite{Gros2015aem,Gros2015,Gros2014Wamot}. However, the metasurface used in \cite{Wanderlinder2017} is not reconfigurable. Consequently, unlike our ERMs, it cannot be used  to simultaneously stir the EM field and improve the field uniformity. 
Finally, we note that the \ecdf\,of the experimental data are increasingly close to the Hill-Ericson-Schroeder regime (dotted purple curves in Fig~\ref{fig:spa}) as the modal overlap increases. Nevertheless, because of the large size of the statistical uncorrelated sample ($\sim 5\times 10^5$ transmission parameters per frequency window \footnote{Note that having an equivalent size of an uncorrelated statistical sample with mechanical stirring is much more difficult. Generally, the size of the statistical ensemble is between one and two orders of magnitude smaller. \cite{Lemoine2007,Lemoine2007b}}) obtained by modulating the boundary condition of the cavity with ERMs, one can still discriminate  between the RMT prediction and the Hill-Ericson-Schroeder regime --- mainly on the tail of the distribution \footnote{The difference between both distributions is also visible on the bulk of the probability distribution function (see insets of Fig~\ref{fig:spa})}. This is the case even for the largest modal overlap regime with $d\sim 20$ studied here (Fig.~\ref{fig:spa}.e).

In conclusion, we experimentally showed that random modulations of a regularly shaped cavity's boundary conditions with simple metasurfaces constitute a new approach to construct a chaotic reverberation chamber without mechanical modifications. Here, we have demonstrated that this approach enables the observation of chaotic behavior for a wide range of modal overlap regimes, even at frequencies as low as the $100^\textrm{th}$ cavity mode. From a practical point of view, in a forthcoming publication \cite{SMM_stirrer}, we will demonstrate how the metasurfaces can create a large number of uncorrelated cavity configurations which is an important features for many applications \cite{Lemoine2007,Lemoine2007b,Krauthauser2005,Lunden2000,Gradoni2013,Sleasman2016,wbac,TondoYoya2017,localiz}. From a more fundamental point of view, these reconfigurable chaotic cavities could be used to verify recent RMT predictions \cite{Savin2017,Fyodorov2017} due to the tantamount realizations they can produce easily and rapidly.

\bigskip

\begin{acknowledgments} 
	The authors thank Olivier Legrand, Ulrich Kuhl and Fabrice Mortessagne from the Universit\'{e} C\^{o}te d'Azur for fruitful discussions and  acknowledge funding from the French ``Minist\`{e}re des Arm\'{e}es, Direction G\'{e}n\'{e}rale de l'Armement''.
\end{acknowledgments}


%
 
 \newpage
 
\onecolumngrid
\section{Supplemental Material: \\ Wave Chaos in a Cavity of Regular Geometry with Tunable Boundaries}
 \subsection{Universal versus non universal behaviors }
 
\begin{figure}[h!]
 	\includegraphics{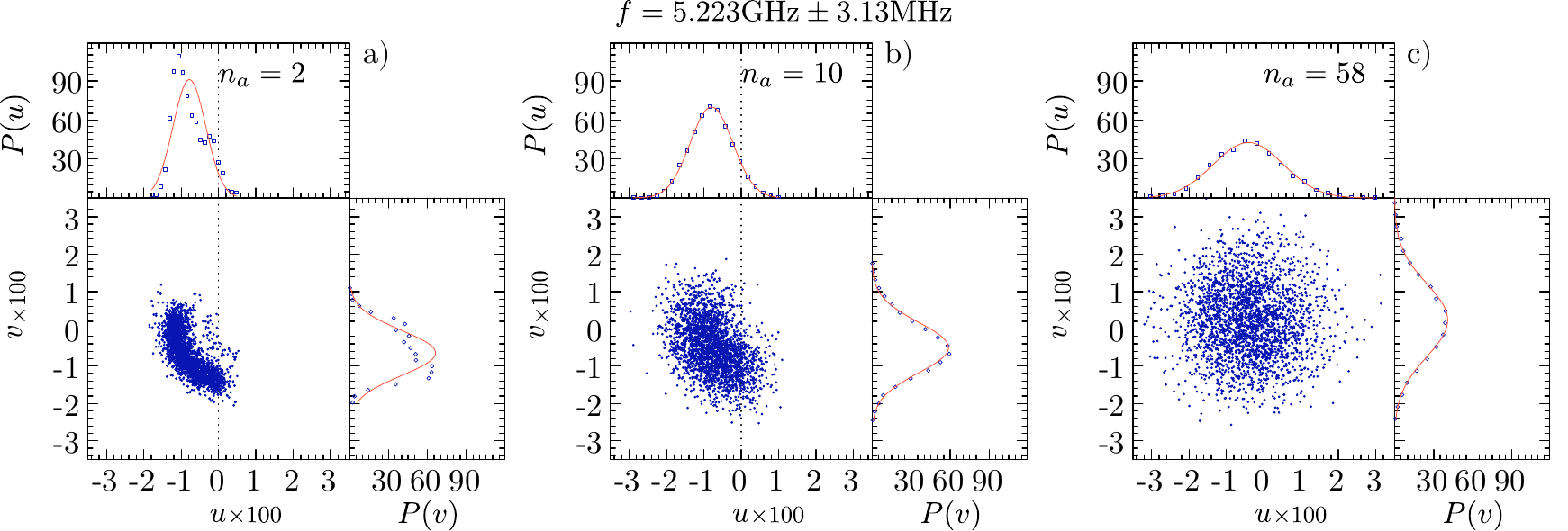}
 	\caption{\label{fig:complement}For different numbers of activated pixels, $n_a$, we plot in the complex plane  the distribution of $S_{12}(f)$ in a small  frequency window of $3.13$~MHz around $f=5.223$~GHz  and for the 500 random configurations of the ERMs. The $\square$ and $\diamond$ respectively correspond  to the marginal distributions functions $P(u)$ and $P(v)$ where  $u=\operatorname{Re}[S_{12}]$ and $v=\operatorname{Im}[S_{12}]$. Each distribution is  compared with the Normal distributions $\mathcal{N}(\mu,\sigma^2)$ (continuous red curve)  where $\mu$ and $\sigma^2$ are the measured mean value and variance  of $u$ and $v$. Cases b) and c) agree with the universal behavior of chaotic cavities. }
 \end{figure}

\begin{figure}[h!]
 	\includegraphics{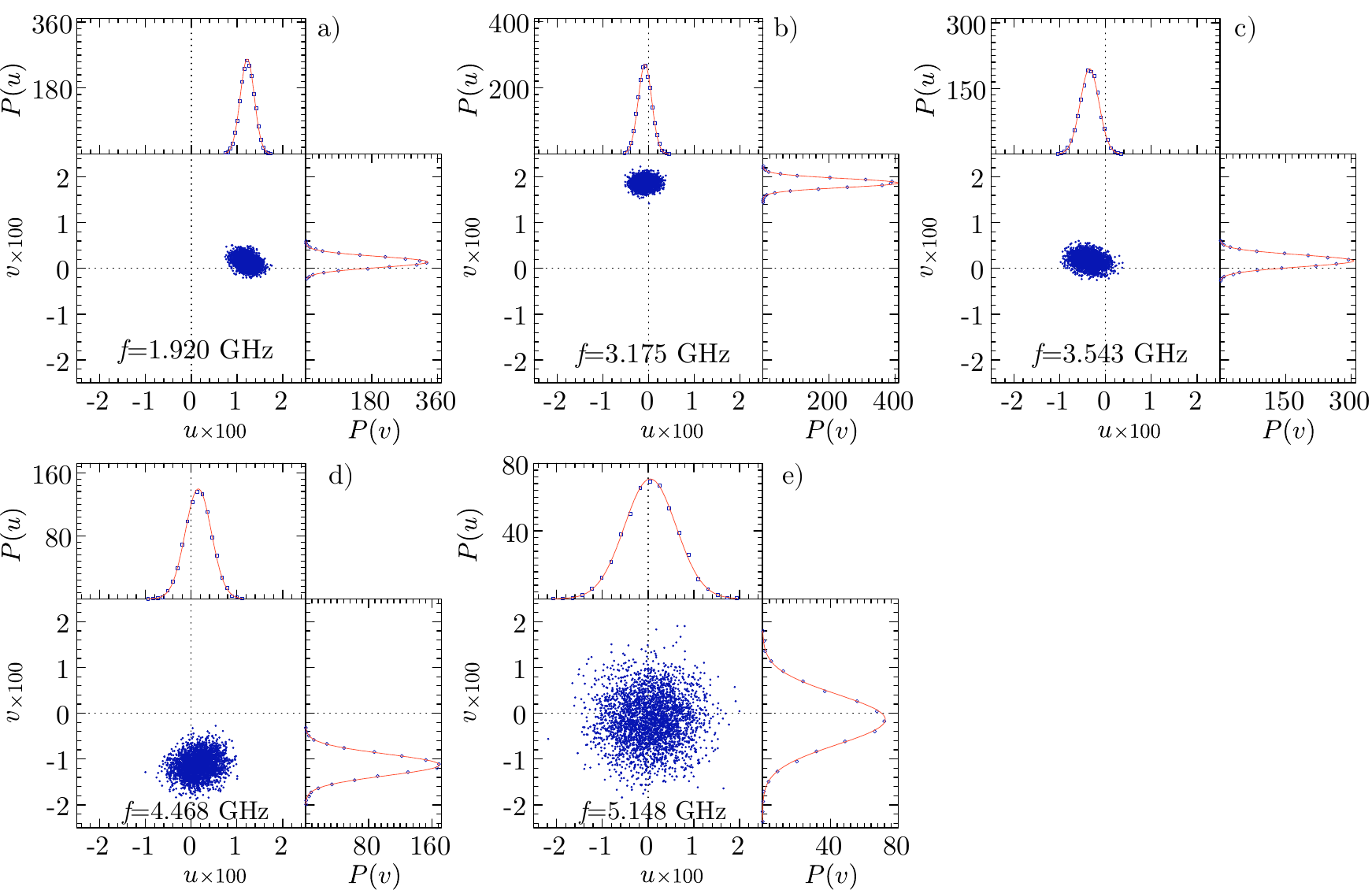}
 	\caption{\label{fig:complement3}Distribution in the complex plane of $S_{12}(f)$ at different fixed frequencies $f$  and for   random 9000 configurations of the ERMs. The $\square$ and $\diamond$ respectively correspond  to the marginal distributions functions $P(u)$ and $P(v)$ where  $u=\operatorname{Re}[S_{12}]$ and $v=\operatorname{Im}[S_{12}]$. $P(u)$ ($P(v)$) is  compared with the Normal distributions $\mathcal{N}(\mu,\sigma^2)$ (continuous red curve)  where $\mu$ and $\sigma^2$ are the measured mean value and variance  of $u$ ($v$). All the cases agree with the universal behavior of chaotic cavities.}
 \end{figure}

 In a chaotic cavity, the only universal statistical requirement for the field is that its real and imaginary parts are Normally distributed. Therefore, this property can be used as an  indicator of the efficiency of the reconfigurable metasurface to make the cavity chaotic.   As illustrated in the Fig~\ref{fig:complement} and Fig~\ref{fig:complement3}, the Gaussianity of real and imaginary parts of the field is systemically verified in our experiments each time the metasurfaces configurations sufficiently impact the boundary conditions of the regular cavity.  Indeed, only the case shown in Fig~\ref{fig:complement}.a), corresponding to frequencies inside the design operation band of the metasurface but an ensemble of configurations where the number of activated pixels of the metasurface are to small (only two), does not agree with the universal behavior of a chaotic cavity. 
 
\begin{figure}[h!]
 	\includegraphics{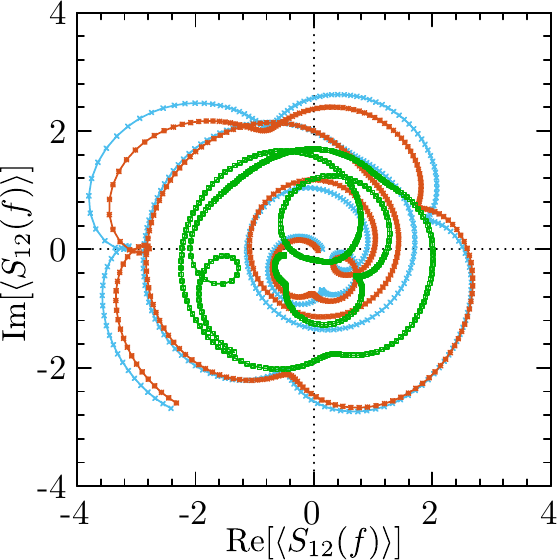}
 	\caption{\label{fig:complement2} For different number of activated pixel, $n_a$, and different frequencies $f$  in a frequency windows of $150$~MHz around  $5.2$~GHz,  we plot the distribution in the complex plane of $\av{S_{12}(f)}$ where $\av{\cdots}$ hold for average on configurations of the ERMs. Blue $\times$, orange $*$, and green $\square$ respectively correspond  with the cases $n_a=2$, $n_a=10$ and $n_a=58$}
 \end{figure}

 The field in chaotic cavities can also display some non universal statistics, which are not related to the chaotic nature of the cavity but are experiment-dependent features. For instance, we observed non vanishing mean values of the complex field (see  Fig~\ref{fig:complement}, Fig~\ref{fig:complement2} and Fig~\ref{fig:complement3}).  In our experiments the latter can be explained by  non negligible direct processes (short path effects) which mainly stem from the directivity and relative antenna positions.  This is illustrated in Fig~\ref{fig:complement3} where the mean values of the complex fields over metasurface configurations follow continuous  and rotating trajectory in the complex plane when the frequency increases.

 \newpage

 \subsection{Random matrix prediction of the normalized field intensity of a chaotic cavity }
 
 We briefly  recall here the main steps leading to this prediction (for details, interested readers can refer to \cite{Gros2016,Gros2015,CRC_review}).
 In presence of losses, for a given configuration of an ideally chaotic cavity (or a given frequency, relying on ergodicity), the real and imaginary parts of each Cartesian component of the field are independently Gaussian distributed, but with different variances \cite{Gros2016,Gros2014Wamot}. The ensuing distribution of the modulus square of each component $\abs{E_a}^2$ depends on a single parameter $\rho$, called the \emph{phase rigidity}, defined by \cite{Gros2016}:
 \begin{equation}
 \rho = \frac{\int_V \E \cdot \E \, d\vr}{\int_V\norme{\E}^2\, d\vr}\; .
 \label{def_rho}
 \end{equation} 
 More precisely, in a chaotic RC, due to the ergodicity of the modes contributing to the response, for a given excitation frequency and a given configuration (here ERMs configurations, polarisations and positions of the antennas), the probability distribution of the normalized intensity of the Cartesian component $ I = \abs{E_a} ^ 2 / \av {\abs {E_a} ^ 2}_{\vr} $ depends solely on the  modulus of $\rho$ and is given by \cite{Kim2005,Gros2016}.
 \begin{equation}
 P(I;\rho)=\frac{1}{\sqrt{1-\modulc{\rho}}} \exp\left[ -\frac{I}{1-\modulc{\rho}} \right] \textrm{I}_0\left[\frac{\modul{\rho} I}{1-\modulc{\rho}} \right]\, . \label{eq:Pnini}
 \end{equation}
 with $\textrm{I}_0$ being the modified Bessel function of the first kind.
 This result was originally proposed by Pnini and Shapiro \cite{Pnini1996} to model the statistics of scalar fields in partially open
 chaotic systems. Note that the above distribution continuously interpolates between the two extreme distributions, namely
 Porter-Thomas for lossless closed systems ($\modul{\rho} \rightarrow 1$) and   exponential
 for completely open systems ($\modul{\rho}=0$ ). The latter case corresponds to the limit where the field is statistically equivalent to a random superposition of traveling plane waves \cite{Pnini1996,Gros2016} meaning that real and imaginary parts of each Cartesian components of the field are statistically independent and identically distributed following a normal distribution. This regime is know as the Hill's regime in the EMC community \cite{Lemoine2007b,hill2009electromagnetic} , the Ericson's regime  in nuclear physics \cite{Dietz2010} or Schroeder's regime in room acoustics \cite{Mortessagne1993} and corresponds to a very high modal overlap regime. 
 Since the phase rigidity is itself a distributed quantity, the distribution of the normalized field intensity  in a chaotic reverberation chamber for an ensemble of responses  resulting from stirring reads
 \begin{align}
 P_I(I)&=\int_0^1  P_\rho (\rho) P(I;\rho) d\rho \label{eq:P:ensemble:rep}
 \end{align}
 where $P_\rho (\rho)$ is the distribution of the phase rigidity of the responses.  Preliminary investigations, based on numerical simulations of the Random Matrix model described in \cite{Gros2016}, show that $P_\rho (\rho)$  depends only on the mean modal overlap $d$.  An Antsatz was proposed in \cite{Gros2015}  to determine $P_\rho (\rho)$ from the only knowledge of $d$. This Ansatz reads: 
 \begin{equation}
 P^W_\rho (\rho) = \frac{2B \exp [-2B \rho/(1-\rho)]}{(1-\rho)^2}\; ,
 \label{Wigner_p_rho}
 \end{equation}
 where the parameter $B$ has a smooth $d$-dependence \cite{Gros2015} numerically deduced from our RMT model presented in \cite{Gros2016}. Originally in \cite{Gros2015} , the empirical estimation of $B(d)$ was  limited to $d\leq 1$. Currently, $B(d)$ have been extended to larger values of $d$ and is given by \cite{CRC_review} 
 \begin{equation}
 B(d) = \frac{a d^2}{1+ bd+cd^2}\, .
 \label{B_de_d}
 \end{equation}
 with   $a= 0.50 \pm 0.02$,  $b=1.35\pm 0.03$  and $c=0.30 \pm 0.02$ \cite{CRC_review}.

\end{document}